\documentclass[letterpaper,twocolumn,10pt]{article}
\usepackage{usenix,epsfig, endnotes}
\usepackage{amsmath,amssymb,amsfonts}
\usepackage{algorithmic}
\usepackage{graphicx}
\usepackage{textcomp}
\usepackage{xcolor}

\usepackage{float}
\usepackage{url}

\usepackage{listings}
\usepackage{hyperref}
\usepackage{cleveref}
\usepackage{biblatex} 

\definecolor{codegreen}{rgb}{0,0.6,0}
\definecolor{codegray}{rgb}{0.5,0.5,0.5}
\definecolor{codepurple}{rgb}{0.58,0,0.82}
\lstdefinestyle{long_code}{
    commentstyle=\color{codegreen},
    keywordstyle=\color{magenta},
    numberstyle=\tiny\color{codegray},
    stringstyle=\color{codepurple},
    basicstyle=\ttfamily\footnotesize,
    breakatwhitespace=false,
    breaklines=true,
    captionpos=b,
    keepspaces=true,
    numbers=left,
    numbersep=5pt,
    showspaces=false,
    showstringspaces=false,
    showtabs=false,
    tabsize=2,
    frame=single,
}
\lstdefinestyle{short_code}{
    basicstyle=\ttfamily\small,
}

\usepackage{CJKutf8} % This is cn support package, common me before print
\begin{document}

%don't want date printed
\date{}

% Make title bold and 14 pt font (Latex default is non-bold, 16 pt)
\title{\LARGE \bf Exploration and Exploitation of Hidden PMU Events}

\author{
{\rm Yihao Yang}$^1$,
{\rm Pengfei Qiu}$^1$,
{\rm Chunlu Wang}$^2$,
{\rm Yu Jin}$^3$,\\
{\rm Dongsheng Wang}$^4$,
{\rm Gang Qu}$^5$\\
$^{1,2,3}$Key Laboratory of Trustworthy Distributed Computing and Service (BUPT), Ministry of Education\\
$^{4}$Tsinghua University~~
$^{5}$University of Maryland~~\\
{\rm khaosyg@gmail.com}, {\rm \{qpf,wangcl\}@bupt.edu.cn},{\rm lambda.jinyu@gmail.com}, \\
{\rm wds@tsinghua.edu.cn}, {\rm gangqu@umd.edu}
}

\maketitle

% Use the following at camera-ready time to suppress page numbers.
% Comment it out when you first submit the paper for review.
\thispagestyle{empty}

\subsection*{Abstract}
Performance Monitoring Unit (PMU) is a common hardware module in Intel CPUs. It can be used to record various CPU behaviors therefore it is often used for performance analysis and optimization. Of the 65536 event spaces, Intel has officially published only 200 or so. \par
In this paper, we design a hidden PMU event collection method. And we found a large number of undocumented PMU events in CPUs of Skylake, Kabylake, and Alderlake microarchitectures. We further demonstrate the existence of these events by using them for transient execution attack detection and build-side channel attacks. This also implies that these hidden PMU events have huge exploitation potential and security threats.

\section{Introduction}
\label{instruction}
Hardware Performance Counter (HPC) is a popular hardware monitoring tool in today's computer architectures. It has been widely used for more than a decade, and these counters can be used to measure events at the CPU level at a granular level, such as instruction execution, Cache Hit or Miss, branch prediction, etc. HPC is very important for performance analysis, code debugging, and optimization. Most modern processor vendors provide HPC support for their processors. In Intel processors, the functional unit used to support HPC is called the Performance Monitor Unit(PMU) \cite{IntelManual}.\par

The Intel official documented more than 200 PMU Events \cite{intel2022perfmon}(which vary slightly on different architectures) for developers to use. These events are introduced initially for code debugging and performance improvement. However, because it can measure various microarchitectural events at a fine-grained level, it has been widely used in various fields, such as malware detection and defense\cite{demme2013feasibility,wang2013numchecker}, microarchitectural attack detection\cite{liCongMiao2019detectSpecRowha,liCongMiao2021detectSpec}, reverse engineering \cite{maurice2015reverseEngineering}, and so on. In addition, some evaluation tools have been designed using PMU for different environment settings, such as PAPI\cite{mucci1999papi}, perf\_event\cite{weaver2013linux},and VTune\cite{Intel2023Vtune}. These tools also greatly facilitate software developers to analyze the performance of their code. In addition to being used for positive work, Qiu et al.\cite{qiu2022pmuspill} found that PMU counters record behaviors of transient instructions besides those of truly committed instructions. They exploited this feature to create a PMU side channel that replicated the Foreshadow attack and compromised the security of Intel SGX\cite{intel2018sgx}.\par

\begin{figure}
    \centering
    \includegraphics[width=0.45\textwidth]{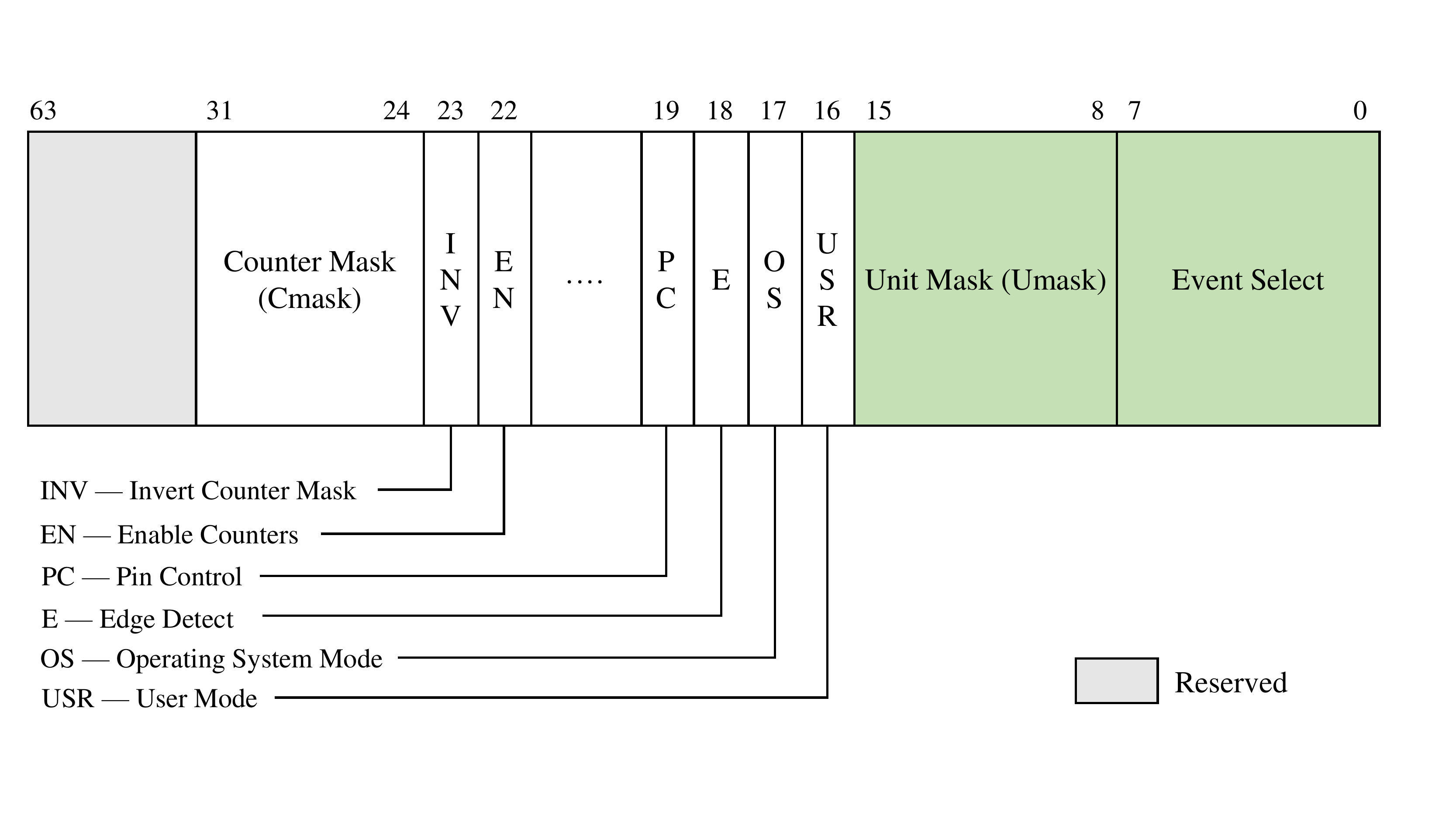}
    \caption{Layout of IA32\_PERFEVTSELx MSRs}
    \label{fig:layout of IA32_PERFEVTSELx MSRs}
\end{figure}

Intel provides 16 bits for PMU event selection\cite{IntelManual}, as shown in Figure \ref{fig:layout of IA32_PERFEVTSELx MSRs}. The 8 bits Umask and 8 bits Event Select respectively make up a complete PMU event. This means that the entire event selection space is $2^{16}$, but even on Intel's latest Alderlake architecture CPUs, only over 200 publicly available Core PMU events exist\cite{intel2022perfmon}. This is a very small subset of the entire event space, meaning there may be many undocumented PMU events. Zhao et al.\cite{zhao2022binoculars}, in their performance analysis and reverse engineering of their work, mentioned two unrecorded PMU events, which they named \texttt{L1D.READ\_REQS} and \texttt{L1D\_BLOCKS.FALSE\_DEPS} based on their event behavior. However, they did not go further. Nick Gregory et al.\cite{gregory2021using} traversed the $2^{16}$ space and filtered for Spectre-sensitive events, and they eventually came up with 81 unrecorded PMU events. But in our work, we traversed the x86 instruction set and recorded the PMU events they triggered. In the end, tens of thousands of undocumented PMU events were found on different microarchitectures, which is far more than 81.\par

In this paper, we design a hidden PMU event collection method and we use the hidden PMU events for transient execution attack detection and construction of side-channel attacks. First, we traverse the x86 instructions using the uops.info\cite{abel2019uops} dataset. At the same time, the monitor program continuously polls the entire PMU event space to collect all possible PMU Events. We found about 20,000 undocumented PMU Events on the i7-6700, and i7-7700 and about 12,000 on the i9-13900k. Then, we try to perform transient attack detection with each PMU event, including Meltdown, Spectre, and Zombieload. Finally, we further demonstrate the effectiveness of these events by constructing side channels for each hidden PMU event and reproducing the above transient execution attacks\cite{spectressb,kocher2020spectre,Lipp2018meltdown,schwarz2019zombieload}.

In general, this paper has the following contributions:
\begin{itemize}
    \setlength{\itemsep}{0pt}
    \setlength{\parsep}{0pt}
    \setlength{\parskip}{0pt}
    \item We have designed a method to collect hidden PMU events. And we found a lot of hidden PMU events on Skylake, Kabylake, and Alderlake.
    \item We performed detection screening for hidden PMU events. We found 377 hidden PMU events that can be used for Meltdown-Detection, 6346 for Spectre-Detection, and 3837 for Zombieload-Detection.
    \item We use these hidden PMU events to build side channels for microarchitecture attacks to further demonstrate their effectiveness. We find 357 hidden PMU events that can be used to build side channels for Meltdown attacks, and 1094 for Spectre attacks.
    
\end{itemize}

These hidden PMU events, similarly, can record some microarchitectural behavior. They can also be used for malware detection, reverse engineering, or even for attack scenarios. Therefore it is important to prove the validity of undocumented PMU events and to assess their hidden security risks.
\section{Background}
\subsection{Performance Monitor Unit}
The Performance Monitoring Unit is an important hardware module on today's processors. It contains a set of performance counters that record various hardware performance events that occur at the CPU level during system runtime. Intel divides the hardware events supported by its performance counters into architectural performance events and non-architectural performance events, which also serve as microarchitectural events\cite{IntelManual}. Architectural performance events refer to events that have consistent behavior across processor architectures, such as Instruction retired, Unhalted core cycles, Branch instructions, etc. Non-architectural performance events are processor microarchitecture specific and have different behavior across different microarchitectures and may vary with processor enhancements. For non-architectural performance events, they are further classified as Core Events, and Uncore Events\cite{IntelManual}. Core Events are defined as performance events that occur inside the CPU, such as Instruction retired, Cache hit or miss, branch prediction, etc. Uncore Events are events that occur in components outside the CPU core, such as memory accesses, I/O operations, and so on. In this article, we will only discuss Core Events.\par

Intel provides users with three fixed counters and four programmable counters\cite{das2019sok, IntelManual}. The fixed counters always monitor fixed events such as logical cycles, reference cycles, etc. The programmable counters are supported by a set of one-to-one event selection MSRs (\texttt{IA32\_PERFEVTSELx}) and performance count MSRs (\texttt{IA32\_PMCx}). The \texttt{IA32\_PERFEVTSELx} MSRs start at address 186H and occupy a contiguous block of MSR address space. Each \texttt{IA32\_PERFEVTSELx} register starting at this address corresponds to an \texttt{IA32\_PMCx} register to start at 0C1H. Intel provides two ways to get the value of the performance counter: Polling or Processor Event-Based Sampling (PEBS)\cite{das2019sok, IntelManual}.
\paragraph{Polling:}The user selects the specified event by changing the value of \texttt{IA32\_PERFEVTSELx} and then reads from \texttt{IA32\_PMCx} the number of times the event occurred. For this purpose, Intel provides specific instructions (\texttt{RDMSR}, \texttt{WRMSR}) to do reads and writes to the MSR.\par
\paragraph{PEBS:}This is a sampling method based on the Performance Monitoring Interrupt(PMI) interrupt. \texttt{IA32\_PEBS\_ENABLE} provides 4 bits of data indicating which \texttt{IA32\_PMCx} overflow condition to enable will trigger the PMI, resulting in the capture of the PEBS record.\par

\subsection{Side Channel Attacks}
\paragraph{Side Channel Attacks:}There are many shared resources in the microarchitecture, such as Cache, TLB, execution ports, etc. The attacker accesses the victim's information by monitoring the state changes of such shared resources. By Side Channel, the attacker does not directly attack the target data but infer the secret information such as the victim encryption key by analyzing the side information (e.g., voltage frequency change, cache timing, etc.) that the microarchitecture inadvertently leaks.\par
In microarchitecture, the most common ones are Cache side channel attacks, such as Flush+Reload\cite{flushreload}, Prime+Probe\cite{liu2015last_prime_probe}, CacheBleed\cite{yarom2017cachebleed}, etc. There are also side-channel attacks based on other shared resources, such as TLBLeed\cite{gras2018translation_tlbleed}, PortSmash\cite{aldaya2019port_portsmash}, Binoculars\cite{zhao2022binoculars} etc. The basic principle of these attacks mostly relies on cache time differences. Qiu et al.\cite{qiu2022pmuspill} established a PMU-based side channel. the PMU captures and records various microarchitectural states, so the victim information can be inferred by analyzing the PMU event counts.

\subsection{Transietn Execution Attacks}
Transient execution attacks are caused by various aggressive optimization strategies introduced by modern processors to improve performance, such as Out-of-Order Execution, Branch Prediction, etc. These strategies may lead to the execution of instructions that should not be executed, which is called transient execution. Although transient instructions are not explicitly committed, they may have some impact on the microarchitecture state. The attacker captures such microarchitecture state changes by establishing side channels and thus inferring the victim's private data. Typical transient execution attacks are Meltdown\cite{Lipp2018meltdown}, Spectre\cite{spectressb,kocher2020spectre}, Zombieload\cite{schwarz2019zombieload}, etc.

\section{Hidden PMU Collector}
\begin{table*}[!ht]
    \centering
    \setlength\abovecaptionskip{5pt}
    \setlength{\belowcaptionskip}{3pt}
    \setlength\arrayrulewidth{1.0pt}
    \caption{Hidden PMU Collector Result~}
    
    \label{tab:CollectorRes}
    \begin{tabular}{rrrrr}
    \hline
        Micro-Architecture & CPU & Total Instructions &  Execution Success & Hidden PMU Events \\ 
    \hline
        Skylake & i7-6700 & 5492 & 3412 & 20599 \\
    \hline
        Kabylake & i7-7700 & 5492 & 3574 & 20230 \\
    \hline
        Alderlake & i9-13900k & 5492 & 3628 & 12503 \\    
       
    \hline
    \end{tabular}
\end{table*}
\begin{figure*}[ht]
    \centering
    \begin{minipage}[b]{1\textwidth}
        \includegraphics[width=1\columnwidth]{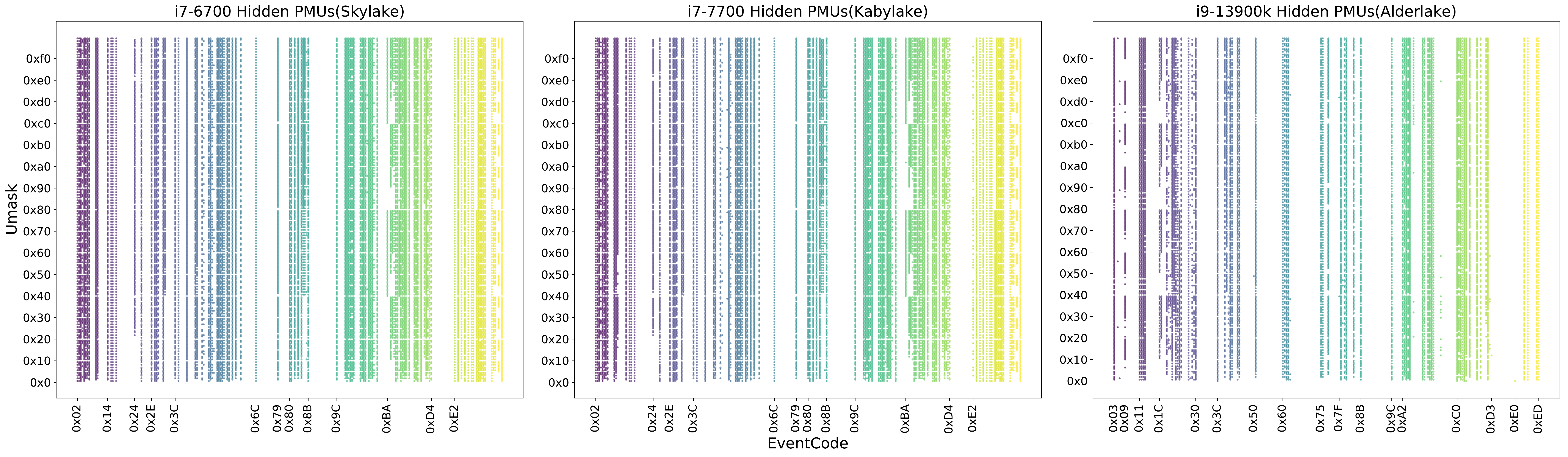}
    \end{minipage} 
    \caption{Distribution of Umaks and EventCode for Hidden PMU Events on different Microarchitectures}
    \label{fig:collector-res}
\end{figure*}
\subsection{Motivation}
As we described in Section \ref{instruction}, PMUs can capture specified types of CPU hardware events to enable developers to optimize their code by understanding the system's runtime characteristics and performance bottlenecks. Today PMUs are widely used in various work scenarios. However, these events represent only a small fraction of the overall event space, and it is worth investigating whether most of the undocumented PMU events can also be used in these scenarios. In addition, PMU is also useful for all kinds of reverse engineering, and it is worthwhile to pay attention to whether Intel's undisclosed PMU implies the existence of some unrevealed CPU hardware components.\par
In addition, there are also some security risks because of the granularity of PMUs. An attacker could detect processor data and instruction flows through the PMU, or create side channels to leak information. Therefore, for a large number of unknown PMUs, their security risks may be even more threatening. Therefore, it is necessary to dig and analyze the hidden PMU events both in terms of positive and negative work.\par
\subsection{Challenges}
\paragraph{x86 Instructions Traversal:}In order to collect hidden PMU events, we tried to execute all x86 instructions to trigger as many PMU events as possible. However, the complexity of the x86 instructions posed a significant challenge to us. We have compiled 5492 instructions based on the uops.info\cite{abel2019uops} dataset, which have different behaviors depending on the processor mode and privilege level. In addition, various jump instructions may cause the program to dead-end or terminate requiring special handling as well. It is worth noting that x86 has evolved with many instruction set extensions requiring specific floating-point units and registers, which different CPUs may support differently.\par
Secondly, the Intel assembly syntax is different from the GCC inline assembly syntax (AT\&T), so we also need to preprocess these instructions. As well, there are multiple types of operands for the same instruction, which in turn increases the complexity of the entire instruction set. Overall, the complexity and diversity of the x86 instruction set have caused a great many problems for us.

\paragraph{Non-deterministic:}As we explained in Section 2, the PMU can monitor various CPU-level events at a fine-grained level. Weaver et al.\cite{weaver2013Non-determinism} show that PMU counting is inherently non-deterministic and over-counting, due to its architectural design. Such uncertainty makes it difficult to determine the validity of hidden PMU events with counts close to zero when collecting PMU events.\par

For such non-deterministic, Das et al.\cite{das2019sok} point out that not all PMU-based work is affected. Among them, malware defense and detection works are more likely to be affected. This is because they rely on the small impact of the attack model on the hardware to determine whether it is being attacked. So, we filter this uncertainty by constructing microarchitectural attack detection models and side channel models with these hidden PMU Events.

\subsection{Hidden PMU Collect}
First, we process the uops.info data set. Because the Intel x86 assembly syntax differs from the GCC inline assembly syntax (AT\&T) in some ways, such as the location of operands, and the register representation. At the same time, we try to keep the registers used by these instructions in a limited range as much as possible, which is convenient for us to fill the operands. It is notable that for some specific extensions, specific registers may need to be used. For this reason, we need to adapt them according to the instruction extensions supported by the CPU. Furthermore, for various jump instructions, we must put the jump target position after the jump instruction, otherwise, it may cause the program to enter a dead loop. Finally, we compiled a list of 5488 instructions.\par
Since we don't know the details of the instruction execution, we need to handle all the exceptions that may arise during the instruction execution. The best way to achieve this is to use Intel Transactional Synchronization Extensions (TSX) to suppress exceptions, which is fast and efficient. Unfortunately, because the success rate of transient execution attacks can be greatly improved with TSXs, many new Intel CPUs do not support this extension. Therefore, we bind all exception signals to custom exception handlers to prevent program crashes. We then fill a limited number of registers with the appropriate values or addresses to adapt the operand types of the instructions.\par
Finally, we monitor the count changes in the 65536 event space before and after each instruction execution and record the readable events and the instructions that triggered them.\par
\subsection{Result Analysis}
\label{umask bit}
We separately performed collection experiments on three machines, as shown in Table \ref{tab:CollectorRes}. Finally, we successfully executed 3412 instructions on the i7-6700 (Skylake) and collected 20599 hidden PMU events. On the i7-7700 (Kabylake), we successfully executed 3574 instructions and collected 20230 hidden PMU events. On the i9-13900k (Alderlake), 3628 instructions were successfully executed and 12503 hidden PMU events were collected.\par
However, we do not think that each of these PMU events corresponds to a microarchitectural behavior. For the EventCode, we found that it is not continuous, this may mean that these events actually exist. In the case of Umask, its distribution makes us wonder if the bit at the specified location determines the event selection condition. As we can see in Figure \ref{fig:collector-res}, the distribution of Umask has a segmented regularity in either microarchitecture. Moreover, some Umask values appear to be invalid under the three microarchitectures mentioned above. As an example, the hidden PMU event with EventCode is \texttt{0x6C} in the i7-6700 shows a certain regular increase in the graph. On further analysis, we find that its Umask values appear to grow as \texttt{0x*1}, \texttt{0x*3}, \texttt{0x*5}, \texttt{0x*7}, \texttt{0x*9}, \texttt{0x*B}, \texttt{0x*D}, \texttt{0x*F}. In binary perspective, the lowest bit of their Umask is \texttt{1}. So we suspect that the value of Umask may be determined by a specific bit.\par
\begin{figure*}[ht]
    \centering
    \begin{minipage}[b]{1\textwidth}
        \includegraphics[width=1\columnwidth]{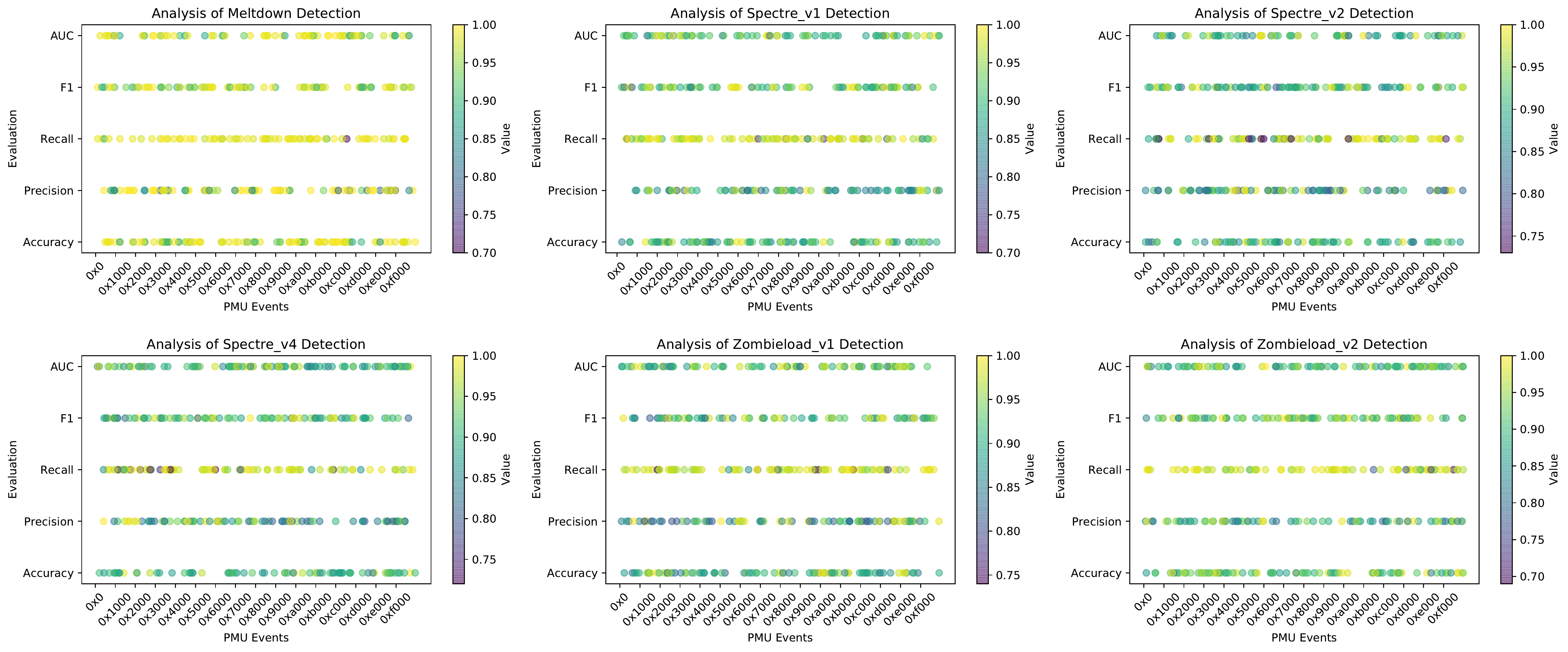}
    \end{minipage} 
    \caption{Transient Execution Attack Detection Model Evaluation}
    \label{fig:detector-res}
\end{figure*}
\section{Application 1:Detecting the Transient Execution Attacks}
To further demonstrate the effectiveness of hidden events, we try to use these hidden events to detect existing transient execution attacks, such as Meltdown\cite{Lipp2018meltdown}, Spectre\cite{spectressb,kocher2020spectre}, Zombieload\cite{schwarz2019zombieload}, etc.
\subsection{Detection Method Design}
In our detection approach, since we do not know the microarchitectural behavior corresponding to each hidden event, we cannot select some specific events for multidimensional detection as in past transient execution detection approaches\cite{liCongMiao2019detectSpecRowha, liCongMiao2021detectSpec}. Instead, we must iterate through each hidden PMU event and monitor their association with these transient execution attacks. To do so, for each attack, we need to collect the count changes for each PMU event in the \texttt{Clean}, \texttt{No-Attack}, and \texttt{Attack} states. The classifier is then trained offline by a machine learning (ML) algorithm and then analyzes the model training results. In this way, we can determine whether that PMU event can be used for that transient execution attack detection.\par
\subsection{Detection Experiment Setup}
\paragraph{Data Collection:}Because most of the transient execution attacks have been fixed on the i9-13900k. So for each hidden event, we collect the count of \texttt{Clean}, \texttt{No-Attack}, and \texttt{Attack} on the i7-6700. The \texttt{Clean} refers to a clean environment where only the victim process is running, to simulate an environment where no malicious attacks exist. In addition, we include text reading and writing to simulate the impact of other third-party processes on PMU counts. For data collection in the \texttt{Attack} environment, we separately have various transient execution attacks running on different logical cores of the same physical core as the victim process, and the monitoring program running on the same logical core as the attacker process, for better data collection. The data collection for each transient execution attack is independent. Here we mainly collect 7 transient execution attacks, namely spectre\_v1\cite{kocher2020spectre}, spectre\_v2\cite{kocher2020spectre}, meltdown(spectre\_v3)\cite{Lipp2018meltdown}, spectre\_v4\cite{spectressb}, zombieload\_v1\cite{schwarz2019zombieload}, and zombieload\_v2\cite{schwarz2019zombieload}.  Finally, to better distinguish the false positives (FP), we need to collect data in the \texttt{No-Attack} environment. In the \texttt{No-Attack} environment, we comment out the Attack Primitive of the attacker process, leaving the rest of the code untouched, and collect the data according to the settings in the \texttt{Attack} environment. It adds false positive(FP) noise to our model so that we can better distinguish it.
\paragraph{Data Processing \& Model Training:}There are many machine learning models used for classification, such as logistic regression (LR), support vector machine (SVM), etc.  Since we need to detect transient execution attacks with 20,000 hidden PMU events respectively, we choose the logistic regression algorithm with less training time. LR is a simple linear classification algorithm that estimates the probability of a given input based on a Sigmoid function. LR algorithm has fewer parameters and less training time compared to other classification algorithms. Moreover, other complex algorithms are theoretically possible if the simple LR algorithm can detect attacks. \par
Then, we label the data. PMU count changes collected in the Attack environment are labeled as 1, and the data in other environments are labeled as 0. This means that 1 indicates that an attack occurred and 0 indicates that no attack. For each type of attack, we collect data in each independent run and use the same number (2000) of samples from both categories to avoid bias. Then, we split the collected dataset into training data (70\%) and test data (30\%). Finally, we train the model with the training data and analyze it with the test data.
\subsection{Experiment Analysis}
In order to evaluate the model better, in addition to the most commonly used \emph{Accuracy} metric, we also calculated other metrics of the detection model, including \emph{Precision}, \emph{Recall}, \emph{F1-Score}, and \emph{AUC} (Area Under Curve). \emph{Precision} represents the proportion of true positive (TP) predicted to be positive, i.e. TP / (TP + FP), which is indicative of false positives (FP). \emph{Recall} represents the proportion of positive samples predicted to be positive, i.e. TP / (TP + FN). \emph{F1-Score} is the average of the two, which takes into consideration both \emph{Precision} and \emph{Recall}. The \emph{AUC} represents the area under the \emph{ROC} (Receiver Operating Characteristic) curve. The \emph{ROC} curve shows the relationship between \emph{Recall} and FP Rate, and the \emph{AUC} is used to measure how well the detection model is able to distinguish between malicious and normal executions. In general, the closer the \emph{AUC} is to 1, the more effective the model is.\par
We screened the detection models with $Accuracy > 0.8, F1 > 0.8, AUC > 0.7 $ and their PMU events Number, and then randomly sampled 400 points to draw a scatter plot as Figure \ref{fig:detector-res}. To prevent model overfitting, we removed the points with index values equal to 1. We also filter out the points with  $F1\in(0.9,1)$ to consider \emph{Precision} and \emph{Recall}. Finally, we got 377 hidden PMU events available for meltdown detection, 530 for spectre\_v1 detection, 4230 for spectre\_v2, 1586 for spectre\_v4, 1823 for zombieload\_v1, and 2014 for zombieload\_v2.\par

\begin{figure}[ht]
    \begin{center}
    \begin{minipage}{0.92\columnwidth}
    
\begin{lstlisting}[language=c,label = {code:gadget}]
zero_pmu();
if(xbegin()==(~0u)){
    asm volatile(
        "cmp (%0), %1"
        "jz equal"
        "nop"
        "jmp end"
        "equal:"
        "   ins1 (eg. movq (%%rax),%%rax)"
        "end:"
        "   ins2 "
        :
        :"r"(The address of Secret),
         "r"(Controllable Variable)
        :
    );
    xend();
}
read_pmu();
\end{lstlisting}
   \label{fig:gadget}
    \caption{Encoding Secret into PMU Side-Channel.}
    \end{minipage}
    \end{center}
\end{figure}

\begin{figure*}[ht]
    \centering
    \begin{minipage}[b]{1\textwidth}
        \includegraphics[width=1\columnwidth]{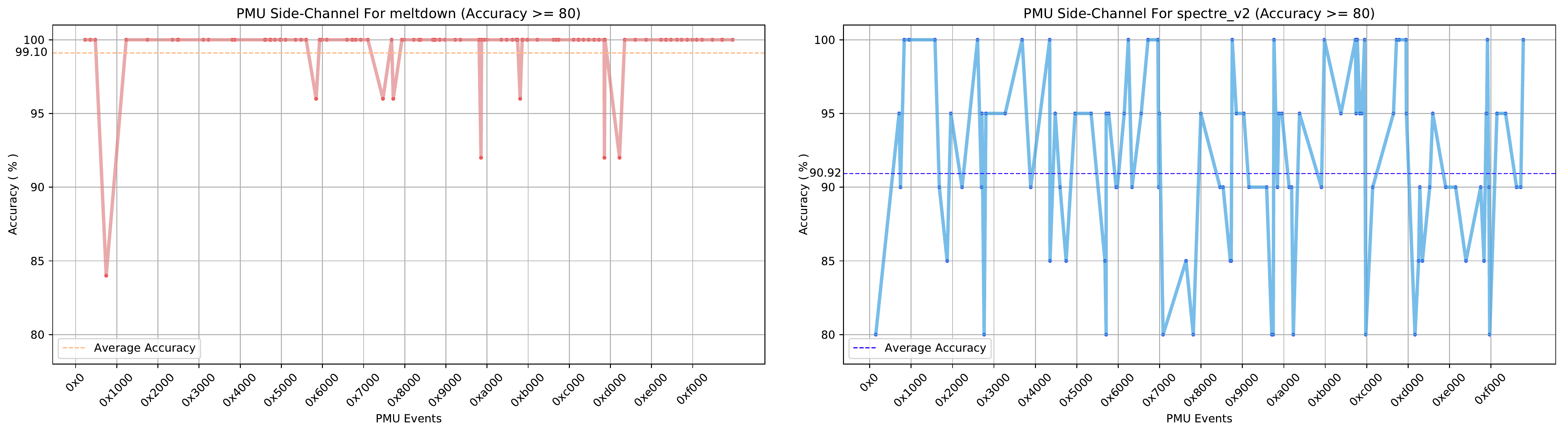}
    \end{minipage} 
    \caption{Accuracy of Hidden PMU Side-Channel Attack}
    \label{fig: Attacker-res}
\end{figure*}
\section{Application 2:Implementing the Side Channel Attacks}
In this section, to demonstrate the potential security threat of hidden PMU events, we attempt to recover private data leaked by transient execution attacks using the hidden PMU event construction side channel.\par
\subsection{Encoding The Secret Data into PMU}
Qiu et al.\cite{qiu2022pmuspill} found that some instructions executed in transient windows also affect the PMU count. Based on this principle they designed an instruction gadget that encodes secret data into the PMU side channel, as in List1. First, the side channel state is cleared, as in the first line of Fig.\ref{fig:gadget}. Then, in the fourth line, we compare the secret data with a controllable variable V, which triggers a transient execution. If the secret data is equal to V, the path of command execution changes, i.e. \texttt{ins1} is executed. And \texttt{ins1} is bound to the PMU event we set, which allows us to infer whether \texttt{ins1} is executed or not from the change in the PMU count, and thus secret data from the controllable variable V.\par
\subsection{The Experiment Setup}
The victim device we chose is also an Intel i7-6700 (Skylake) processor with 32 KiB, 8-way L1 data Cache, on Ubuntu 16.04 with kernel version 4.15.0. We successfully reproduced two transient execution attacks, Meltdown and Spectre\_v2, using the above instruction gadget. We also tried the same for Spectre\_v1, but it did not work. A brief analysis of the reason for this may be the existence of branching instructions in our gadget, which we suspect may affect branch mistraining.\par
Theoretically, the best way is to perform a combined traversal of instruction and PMU Events. However, even if we set \texttt{ins2} to \texttt{nop}, this would require $5492*20599 \approx 1.13*10^8 $ iterations with an average time of 0.4s per iteration, which would take about a year, and this is not acceptable. In addition, we try to record the corresponding instructions that trigger a PMU in the Collector. Then we only traverse the combination of these, which can reduce the iteration space to about 11 million iterations, but this also takes more than 30 days.\par
So, we set \texttt{ins1} to be a single access instruction, because according to the results in Collector, the access instruction can trigger the most PMUs. Then only 20,599 PMU Events are traversed, which eventually reduces the traversal time to about one hour. Although this loses some precision, it also demonstrates the potential security threat of these hidden PMU events.\par
\subsection{Experiment Analysis}
Throughput rate and error rate are two important measures to evaluate side-channel attacks. The throughput rate is mainly determined by the instruction gadget execution time, the number of iterations, the exception handling time, or the branch training time. On our experimental device (i7-6700), the instruction gadget was iterated 10 times to recover the secret data. For the meltdown attack, the average throughput rate can reach 789.86 Bps if Intel TSX exception suppression is used. if the exception signal processing function is used, the throughput rate drops to 497.49 Bps. while for spectre\_v2, the average throughput rate is around 148.68 Bps because the branch training takes longer than the exception processing.\par
Another important metric is the error rate (or accuracy). Unlike the throughput rate, the accuracy of an attack is determined by the individual PMU events. Different PMU events have different Accuracy, so we iterate through 20,000 hidden PMU events and filter out the event numbers with $Accuracy \ge 80$, and then calculate their average accuracy. For demonstration purposes, we randomly sampled 100 samples and show them in Fig. \ref{fig: Attacker-res}. It can be seen that for the meltdown attack, the average error rate($1- Accuracy$) is 0.9\% for 10 iterations and 9.08\% for spectre\_v2. Finally, we obtain 357 hidden PMU events that can be used to construct a side channel to recover the secret data leaked for the meltdown attack with $Accuracy \ge 80$, and 1094 for spectre\_v2.

\section{Discussion \& Future Work}
\subsection{Limitations}
The first is that this paper only tries to explore hidden PMU events in three microarchitectures of Intel CPUs, so whether there are also hidden events in other microarchitectures or there are also undocumented PMU events in other processor vendors' CPUs. Second, this paper does not work reverse for all the hidden events, i.e., find their respective corresponding microarchitectural behaviors. Then, only two transient execution attacks are successfully reproduced using the hidden PMU side channels, and whether other transient execution attacks can also recover private data using hidden PMUs as side channels. Or are there any other security threats for these hidden PMUs other than as side channels? Also, whether these hidden PMU events correspond to some unknown hardware components in the microarchitecture. We leave these questions to be explored in the future.\par
\subsection{Future Extensions}
\paragraph{Reverse-Engineering:}We have found a large number of hidden PMU events and associated these events with various microarchitecture attacks. Although we collected the events by associating the hidden events with the instructions that triggered them, we did not do further analysis of the specific behavior of the instructions. However, because of the enormous number of events, we were unable to match the event codes to the microarchitecture behaviors as Intel officially disclosed events. As we introduced before, a PMU event consists of 16 bits. For the high 8 bits (Umask) most of the possible values appear in our collection of events, while for the low 8 bits (Event Select) it is limited to a fixed range, which also determines the general class of events. Therefore, we believe that it is feasible and necessary to reverse the hidden Event Select Code in future work.\par
\paragraph{Specific bit decision Umask:}As we mentioned above, Umask appears in most of the possible values of the events we collected. However, we do not believe that each Umask represents an event condition, so we suspect that the CPU only focuses on the specified bit of Umask when checking the event number. In this paper, although the pattern of Umask distribution was initially analyzed in Section \ref{umask bit}, it was not further explored. Therefore, it is necessary to explore the principle of Umask selection in future work.\par
\section{Conclusion}
PMU were originally designed for software performance optimization, but because of their granularity of monitoring, they have been widely used in various scenarios. In this paper, we found a large number of undocumented PMU events in microarchitecture CPUs such as Skylake, Alderlake, etc. Based on this discovery, we use these hidden PMU events to accomplish the detection of various transient execution attacks, as well as the leakage by encoding private data into PMU events during transient execution to build side channels.\par
Our experiments show that these hidden PMU events do exist and can be used for positive or negative work. Also at the end of the paper, we analyze the limitations and possible extensions of this study. Future work on hidden PMU events is worth exploring.

{\footnotesize }
\printbibliography

\end{document}